\documentclass[12pt,aps,showkeys,preprint]{revtex4-2}

\usepackage{color}
\usepackage{graphicx}
\usepackage{float}
\usepackage{epsfig}
\usepackage{epstopdf}
\usepackage{subfig}
\usepackage{amsmath} 
\usepackage{amsmath,amssymb,amsfonts} 
\usepackage{bm} 
\usepackage{hyperref}
\usepackage{titlesec}

\titleformat*{\section}{\Large\bfseries}
\titleformat*{\subsection}{\large\bfseries}
\titleformat*{\subsubsection}{\bfseries}
\titleformat*{\paragraph}{\bfseries}
\titleformat*{\subparagraph}{\bfseries}

\def\begeqn{\begin{equation}}
\def\endeqn{\end{equation}}
\def\begeqnar{\begin{eqnarray}}
\def\endeqnar{\end{eqnarray}}
\def\begeqnarn{\begin{eqnarray*}}
\def\endeqnarn{\end{eqnarray*}}

\usepackage{braket} 

\begin{document}
\baselineskip 18pt  

\title{Optimal social security timing}
\date{\today}
\author{A.Y. Aydemir}
\email{aydemir@utexas.edu}

\begin{abstract}
The optimal age that a retiree claims social security retirement benefits is in general a complicated function of many factors. However, if the beneficiary's finances and health are not the constraining factors, it is possible to formally derive mathematical models that maximize a well-defined measure of his total benefits. A model that takes into account various factors such as the increase in the benefits for delayed claims and the penalties for early retirement, the advantages of investing some of the benefits in the financial markets, and the effects of cost-of-living adjustments shows that not waiting until age 70 is almost always the better option. The optimal claiming age that maximizes the total benefits, however, depends on the expected market returns and the rate of cost-of-living adjustments, with the higher market rates in general pushing the optimal age lower. The models presented here can be easily tailored to address the particular circumstances and goals of any individual.
\end{abstract}
\keywords{Social Security; optimal timing}
\maketitle

\section{Introduction}
Social Security retirement benefits are an essential part of the ``Social Safety Net'' in the US and provide much needed income to many retirees. Thus, at what age they are claimed is usually not an option but dictated by financial necessity; a beneficiary files for benefits as soon as or soon after he is qualified. However, there are undoubtably many cases where the beneficiary can afford to wait for an optimal age for claiming his benefits. Optimization criteria can vary greatly, of course, depending on the circumstances and goals of the individual. In general, maximizing some measure of the total lifetime benefits can be an appropriate goal when health issues and financial stress are not important concerns. 

A number of authors over the years have studied the question of the ``optimal age for claiming Social Security benefits'' from various points of view, with differing methodologies. Here we give only a very brief summary and refer the reader to the extensive references in the works cited here. Coile {\em et al.}\cite{coile2002} establish the benefits of delayed claims using financial simulations and find that their results are in general consistent with the behavior observed in various cohort groups. Friedman and Phillips \cite{friedman2008} take what they refer to as a ``sequential approach'' in which a decision to claim or postpone is made each year. Perhaps not surprisingly, they find that a particular decision that is appropriate for a given year may not be the correct one for the following year. Alleva\cite{alleva2016} uses a range of discount rates to calculate the present value of a future benefits stream to find an optimal age for claiming benefits. As expected, higher discount rates lead to a correspondingly lower optimal age. Ali {\em et al.}\cite{ali2019} do a similar optimization of the discounted present value of the future benefits but make use of numerical simulations to take into account various factors such as interest rate and life expectancy risks.
A model based on income and wealth-level changes that includes asset risks and mortality by Diamond {\em et al.}\cite{diamond2021} seems to suggest that delaying until age 70 is usually the better option unless market returns of over $7\%$ can be guaranteed, which differs from our conclusions. However,  we are unable to pinpoint the source of this difference due to an apparent lack of transparency in their methodology. 

The problem of determining an optimal age for claiming benefits, stripped of concerns about longevity, market risks and general economic fluctuations, essentially boils down to an interesting optimization exercise in a multi-dimensional parameter space. In our view, with some simplifying assumptions, this problem can be attacked essentially analytically, yielding expressions immediately useful for most individuals. They can also be used in future stochastic simulations or other studies to explore the effects of various risk factors.

The opportunity for optimization arises from the way the benefits stream is structured. Briefly, the benefits are reduced if claimed before the full retirement age (FRA) at 66 or 67\cite{fra2022}; similarly they increase for delayed retirement, until age 70\cite{delayedRet2022}. However, there are opportunity costs associated with delayed claims, since the benefits claimed early can be invested in the financial markets. These factors all combine to make the task of determining the optimal timing of Social Security retirement benefits a challenging problem, even without  longevity  or market risk concerns. 

Section~\ref{sect:breakEven} is a review of the usual {\em break even} analysis, the simplest calculation that a potential beneficiary can do to determine the age he would break even if he were to delay his claim. This Section also examines how the cost-of-living adjustments (COLAs) would modify these calculations. Section~\ref{sect:marketGains} defines a {\em gain function}, a measure of the relative merits of claiming early versus late (at age 70) and studies how it varies with various parameters such as the claiming age, an assumed average market return rate and COLAs. Section~\ref{sect:optimize} discusses two possible optimization paths for the claiming age. Finally Sect.~\ref{sect:summary} presents a summary and conclusions of this study.

\section{A break-even analysis} \label{sect:breakEven}
Here we introduce some of the parameters and terminology, all in the context of a commonly-performed break-even analysis. Assume that $S_0$ is the yearly benefit amount that would be available at age 70, i.e., the maximum benefit allowed under the Social Security rules for a given prospective retiree. Since the benefits can be claimed as early as age 62, we will always be comparing two cases:
\begin{itemize}
\item{\bf The early scenario:} The beneficiary claims his benefits starting $K$ number of years before age 70, where $1\le K \le 8$, or
\item{\bf The late scenario:} He waits until age $70$ when his benefits reach the maximum amount $S_0$.
\end{itemize}

Presently the social security benefits increase by $8\%$ for every year they are delayed beyond the full retirement age (FRA); the increases stop at age 70\cite{delayedRet2022}. We will use the parameter $p$ to denote this rate of increase, which, unless otherwise noted, will have the value $p=0.08$ throughout this work. The benefits are reduced, but at a smaller rate, if claimed before the full-retirement age\cite{fra2022}. In order to simplify the calculations we will assume symmetry about FRA and use $p$ for the rate of reduction also. This simplification will make our results somewhat more pessimistic, from the point of view of an early claimer, for claims made before FRA.

With this assumption, if the benefits are claimed $K$ years before age 70,  the annual amount is reduced to
\begeqn
S_K = \frac{S_0}{(1+p)^K}, \label{eqn:SK}
\endeqn
where  $p=0.08$. As stated earlier, the comparison is always with the ``late scenario'' where the beneficiary claims at age 70 when his annual benefit amount is $S_0$.

Since $S_0 > S_K$, the total earnings with the late scenario, $T_L$, will eventually exceed those from the early scenario, $T_E$. After $n$ number of years past age 70, the total benefits for the early and late scenarios are
\begeqnar
T_E(K,n,p) & = & (K+n)S_K = \frac{(K+n)S_0}{(1+p)^K}, \text{~and} \label{eqn:TE} \\
T_L(n) & = & nS_0. \label{eqn:TL}
\endeqnar
Note that we find it convenient to count the years before and after age 70 with different variables, $K$ and $n$, respectively. Thus, $K=2$ will refer to age $70-K=68$. Similarly $n=10$ will refer to age $70+n=80$.

Let $n=n_1(K,p)$ be the solution of the equation $T_E = T_L$, the break-even point. We can easily show that
\begeqn
n_1(K,p) = \frac{K}{(1+p)^K - 1},~~K \ge 1. \label{eqn:n1}
\endeqn
A beneficiary who opts for the early scenario $(1\le K \le 8)$ starts falling behind at age $70+n_1$; beyond that point, his cumulative benefits ($T_E$ in Eq.~\ref{eqn:TE}) will be less than $T_L$ that he would have enjoyed if he had waited until age 70. In other words, for given values of the parameters $K,p$, we have
\begeqn
T_E(K,n,p) < T_L(n) \text{~for~} n > n_1. \label{eqn:TEvsTL}
\endeqn
Equivalently, viewed from the point of view of a late claimer, the break-even point comes at age $70+n_1$, after which he gets further ahead every year. (In this work, we will view the results mostly from an early claimer's point of view.)

For some numerical examples, we have (with $p=0.08$):
\begeqnar
K=1 &: & ~n_1 = \frac{1}{p} = 12.5 \text{~years}, \nonumber \\
K=8 &: &  ~n_1 = 9.4 \text{~years.} 
\endeqnar
Thus, someone who claims benefits at age 69 ($K=1$) will start falling behind at age $70+12.5 \simeq 83$.
A more impatient beneficiary who starts at age 62 ($K=8$) will be behind by the time he reaches his $80^{th}$ birthday.
Under these conditions, and excluding all other factors (market gains, poor health, spousal benefits, etc.) it is better to wait until age 70 to claim your benefits, assuming that you will live beyond age 83. Note that this analysis is independent of $S_0$, the benefit amount at age 70, and depends only on $p$, the annual rate of increase in the benefits before age 70. Further results for $1 \le K \le 8$ are shown in the column labeled $n_1(q=0)$ of Table~\ref{table:Kn1CTable}.

Next we examine how these results are modified by the cost-of-living adjustments (COLAs).

\subsection{Effects of the cost-of-living adjustments} \label{sect:noMwithC}
The benefits increase in general every year due to cost-of-living adjustments implemented to compensate for inflation. The data for these annual increases is available for the years 1975-2022 at the Social Security Administration website\cite{cola2022}.

Future COLA's are of course unknown. For the purposes of these calculations we choose to use an average COLA $q$ based on $N$ years of historical data; here the average $q$ is defined by
\begeqn
(1+q)^N = \prod_{i=1}^N(1+q_i), \label{eqn:qDef0}
\endeqn
where $q_i$ is the COLA for the $i^{th}$ year. More explicitly, $q$ can be put in the form
\begeqn
q = \exp\left\{\frac{1}{N}\sum_{i=1}^N\ln(1+q_i)\right\} - 1. \label{eqn:qDef}
\endeqn
Using the full set of data for 1975-2022 in Eq.~\ref{eqn:qDef} results in an average COLA of $q=0.03745$ or $3.7\%$. If we exclude the high-inflation period of the late 1970's and average only over the last 40 years (1983-2022), we find $q=0.02508$ or $2.5\%$. 

Since the cumulative effect of these adjustments can be substantial over time, here we examine how they modify our earlier results, in particular Eqs.~\ref{eqn:SK}-\ref{eqn:n1}. We again assume $S_0$ is the annual benefit amount at age 70. As in the ``early scenario'' above, if the beneficiary claims benefits $K$ years earlier, $S_K$ of Eq.~\ref{eqn:SK} has to be discounted by $K$ number of future COLA increases; thus, the starting year benefits will now be
\begeqn
S_K^C = \frac{S_K}{(1+q)^K} = \frac{S_0}{(1+p)^K(1+q)^K}, \label{eqn:SKC}
\endeqn
where we use the superscript $C$ to denote COLA-adjusted values. Then, $n$ years after age 70, the total benefits received will be
\begeqn
T_E^C = S_K^C\sum_{i=0}^{K+n-1}(1+q)^i=\frac{S_0}{(1+p)^K(1+q)^K}\frac{(1+q)^{K+n}-1}{q}. \label{eqn:TEC}
\endeqn
Since $\lim_{q \rightarrow 0}[(1+q)^{K+n}-1]/q=(K+n)$, $T_E^C$ reduces to $T_E$ of Eq.~\ref{eqn:TE} for $q \to 0$ as expected. In the ``late scenario'' where the benefits start at age 70, the total after $n$ years will now be
\begeqn
T_L^C = S_0\sum_{i=0}^{n-1}(1+q)^i= S_0\frac{(1+q)^n - 1}{q}. \label{eqn:TLC}
\endeqn
Again, $T_L^C$ reduces to $T_L$ of Eq.~\ref{eqn:TL} in the limit $q \rightarrow 0$. In order to find when an early claimer starts to fall behind, for given $K,p,q$, we need to solve the equation
\begeqn
T_E^C(K,n,p,q) = T_L^C(n,q) \label{eqn:TECeqTLC}
\endeqn
for $n$. We can easily show that the solution is (for $q > 0$) $n=n_1$, where
\begeqnar
n_1 & = & \frac{a_1(K,p,q)}{a_2(q)}, \label{eqn:n1C} \\
a_1 & = & \ln\left\{\frac{(1+p)^K(1+q)^K - 1}{(1+q)^K[(1+p)^K - 1]}\right\}, \nonumber \\
a_2 & = & \ln{(1+q)}. \nonumber
\endeqnar

\begin{table}[h!]
\centering
\begin{tabular}{| c ||  c ||  c | c |} 
 \hline
 $~K~$ & $~n_1(q=0)~$ & $~n_1(q=0.025)~$ & $~n_1(q=0.037)~$ \\ [0.5ex] 
 \hline
 1 & 12.5 & 10.78 & 10.15 \\ 
 2 & 12.02 & 10.30 & 9.67  \\
 3 & 11.55 & 9.84 & 9.21 \\
 4 & 11.10 & 9.39 & 8.77  \\
 5 & 10.65 & 8.95 & 8.34 \\
 6 & 10.22 & 8.54 & 7.93 \\
 7 & 9.81 & 8.13 & 7.53 \\
 8 & 9.40 & 7.74 & 7.15 \\ [0.5ex] 
 \hline
\end{tabular}
\caption{\baselineskip 12pt The break-even point $n_1(q)$ for three different values of the average COLA parameter $q$ is shown: $q=0$ (no COLA), $q=2.5\%$ (average for the years 1983-2022), and $q=3.7\%$ (average for the years 1975-2022). $p=0.08$ is assumed. Recall that $n_1$ measures the number of years after age 70.}
\label{table:Kn1CTable}
\end{table}

Note that Eq.~\ref{eqn:n1C} is valid in the limit $q\rightarrow 0$ but is not defined at $q=0$. Recall that $n_1$ is the break-even point: For given $K,p,q$, and early claimer starts falling behind at age $70+n_1$, i.e., $T_E^C(n) < T_L^C(n)$ for $n > n_1$. Table~\ref{table:Kn1CTable} shows that $n_1$ decreases as the COLA rate $q$ increases. For example, if the beneficiary claims at age 66 (K=4), he starts falling behind at age 81 ($n_1=11.1$) with no COLA ($q=0$). If the average COLA is $q=3.7\%$, he falls behind by age 79 ($n_1=8.77$), two years earlier. This general decrease in $n_1$ with COLA is mostly due to the decrease in the initial benefits: $S_K^C = S_K/(1+q)^K < S_K$ (All scenarios assume a fixed yearly benefit $S_0$ at age 70). In summary, taking into account the cost-of-living adjustments makes early claims less attractive. 
Below we investigate another scenario where claiming early may lead to a more positive outcome.

\section{Taking advantage of possible market gains} \label{sect:marketGains}
In the previous section, we assumed that, if the beneficiary claimed his Social Security benefits early (before age 70), he would be using them to meet his financial needs.
The analysis in this section is geared towards those beneficiaries who can afford to wait until age 70 but may nevertheless want to claim earlier, with the intention of investing the payments they receive in the financial markets. Clearly there is no unique way to proceed in this scenario. Investments can be made in myriad different ways. In order to obtain some clear, quantitative results, we assume a rather simple scheme that can be improved with more realistic details at a later date if necessary.

\subsection{``Early claim with market gains'' scenario--no COLAs} \label{sect:withMnoC}
We consider a rather straightforward scheme and, for the time being, we ignore the cost-of-living adjustments. The assumptions we make are as follows:
As before, the beneficiary files for benefits $K$ years before age 70, where $1 \le K \le 8$. But now, at the end of each year, until he is 70, he invests the benefits received that year in a financial instrument with an average annual return rate of $r$. Starting at 70, he stops diverting his benefits to the financial markets but allows the already-accumulated sum to grow indefinitely at the assumed annual rate of $r$.

Because of the exponential growth of the funds in the market (earnings compounded annually), it is possible for this scenario to generate more total income (integrated over some relevant period) than the {\em late scenario} of the previous section. Here we determine a range of values for the parameters $K, r$ for which such a positive outcome is possible.

The average rate of return $r$, averaged over some number of years $N$ (for our purposes, $N \simeq 30$ years) is defined similarly to the COLA parameter $q$ of Eqs.~\ref{eqn:qDef0}, \ref{eqn:qDef}:
\begeqn
r = \exp\left\{\frac{1}{N}\sum_{i=1}^N\ln(1+r_i)\right\} - 1, \label{eqn:rDef}
\endeqn
where $r_i$ is the rate for the $i^{th}$ year.
In order to explore the possible advantages of this scenario, we start with a calculation of the total sum, $T_M$, that the beneficiary accumulates in the market before he reaches age 70, i.e., the total value of his social security investments in the market when he turns 70. Recalling Eq.~\ref{eqn:SK} and using the assumptions outlined above, this sum is given by
\begeqn
T_M  =  S_K\sum_{i=0}^{K-1}(1+r)^i = \frac{S_0}{(1+p)^K}\frac{(1+r)^K - 1}{r},~~K\ge 1. \label{eqn:TM}
\endeqn
By  assumption, this sum will be left in the market to grow; therefore, $n$ years after age 70, the  total benefits for the {\em early claim with market gains scenario} will be
\begeqn
T_E^M = T_M(1+r)^n + nS_K. \label{eqn:TEM0}
\endeqn
The first term on the right-hand side represents the growing amount in the markets. The second term is the total received (and not invested) after age 70. $T_E^M$ can be written explicitly as
\begeqn
T_E^M = \frac{S_0}{(1+p)^K}\left\{\frac{(1+r)^n\left[(1+r)^K-1\right]}{r} + n\right\}. \label{eqn:TEM}
\endeqn
This equation replaces Eq.~\ref{eqn:TE} of the previous section. Note that in the limit $r\rightarrow 0$ (thus removing the market gains), Eq.~\ref{eqn:TEM} reduces to Eq.~\ref{eqn:TE} as expected, since
\begeqn
\lim_{r\rightarrow 0}\frac{(1+r)^n\left[(1+r)^K-1\right]}{r} = K.
\endeqn

At this point it will be useful to define a relative {\em gain function} $g$ that will measure the early claimer's gains (or losses) as a function of time:
\begeqn
g(K,n,p,r) \equiv \frac{T_E^M(K,n,p,r)-T_L(n)}{T_L(n)}, \label{eqn:gainM0}
\endeqn
where $T_L(n)$ is given by Eq.~\ref{eqn:TL}. For fixed $K,p,r$, a positive gain function implies that the beneficiary is ahead ($n$ years after 70) with respect to the ``late scenario'' where he waits until age 70 to start collecting his benefits. 

Using Eqs.~\ref{eqn:TL}, \ref{eqn:TEM}, the gain function can be written explicitly in the form
\begeqn
g(K,n,p,r) = \frac{1}{(1+p)^K}\left\{\left[\frac{(1+r)^K-1}{r}\right]\left[\frac{(1+r)^n}{n}\right] + 1\right\} - 1. \label{eqn:gainM}
\endeqn
For fixed $K,p,r$ we have
\begeqn
\lim_{n \to 0} g(K,n,p,r) = \infty,~~\lim_{n \to \infty} g(K,n,p,r) = \infty.
\endeqn
Thus, $g(K,n,p,r)$ is a non-monotonic, convex function of $n$ with a minimum at some $n=n^\ast$. 
The point where the gain function attains its minimum value (again for fixed $K,p,r$) can be calculated easily. Treating $n$ as a continuous time variable and letting
\begeqn
\left.\frac{\partial g}{\partial n}\right|_{n=n^\ast} = 0
\endeqn
leads to
\begeqn
n^\ast(r) = \frac{1}{\ln{(1+r)}}, \label{eqn:nStar}
\endeqn
which is independent of the parameters $K,p$. Making use of $n^\ast$, we find the minimum gain
\begeqn
g_{min}(K,p,r) = g[K,n^\ast(r),p,r]. \label{eqn:gmin}
\endeqn
In the following analysis, the return rate $r=r^\ast$ for which $g_{min}=0$ will play an important role; it can be calculated by solving the equation
\begeqn
g[K,n^\ast(r^\ast),p,r^\ast]=0 \label{eqn:rStar}
\endeqn
for $r^\ast$ with given $K,p$. In general this step requires a numerical solution, but the case $K=1$ can be treated analytically because of simplifications in the function $g(K,n,p,r)$. For fixed $p$ and $K=1$, Eqs.~\ref{eqn:nStar} and \ref{eqn:rStar} lead to
\begeqn
n^\ast = \frac{e}{p},~~~~r^\ast = e^{p/e}-1~~~~(K=1), \label{eqn:Keq1solns}
\endeqn
where $e$ is the base of the natural logarithm. For $K>1$ the analytic results are not very illuminating, and we use Mathematica\cite{mathematica2022} for numerical solutions. Values of $n^\ast,~r^\ast$ for $1\le K \le 8$ without the COLA modifications are tabulated in the columns labeled $n^\ast(q=0)$ and $r^\ast(q=0)$, respectively, of Table~\ref{table:KnrTable}.

An important property of the gain function that will be useful below is
\begeqn
r_1 > r_2 \Rightarrow g(K,n,p,r_1) > g(K,n,p,r_2), \label{eqn:gCompare}
\endeqn
which follows directly from Eq.~\ref{eqn:gainM}. It is also intuitively obvious, since higher market yields will naturally lead to higher relative gains.

\begin{figure}[htbp]
\vspace{-0pt}
\begin{center}
\includegraphics[width=4in]{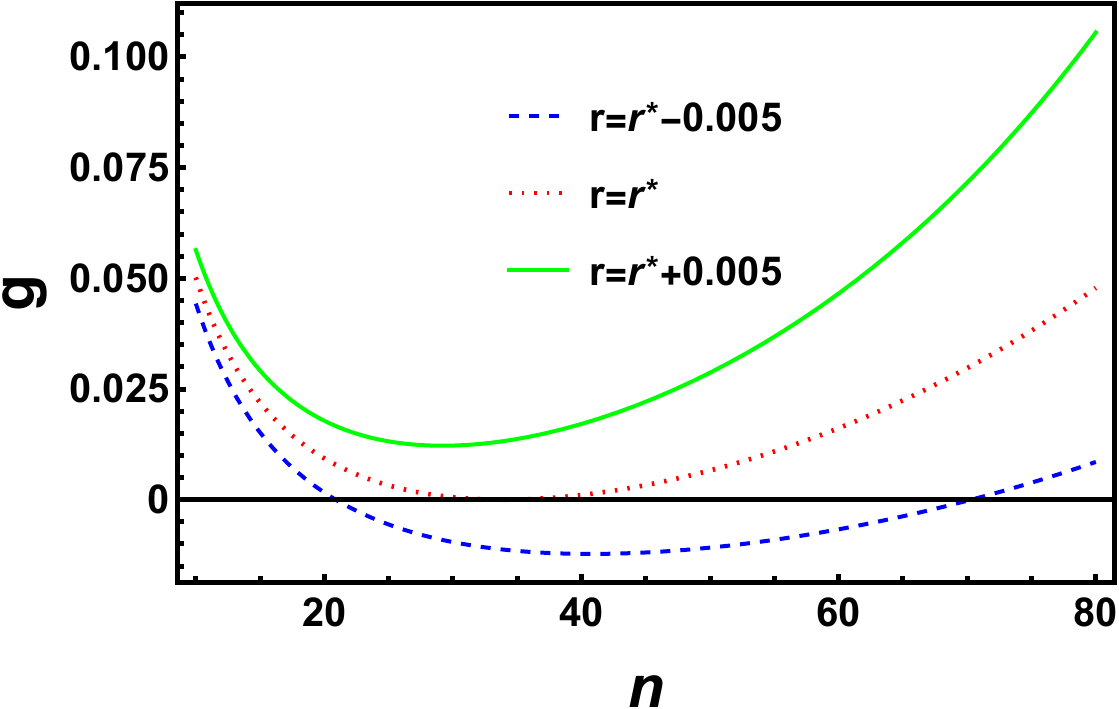}
\caption{\baselineskip 12pt The gain function $g(K,n,p,r)$, with no COLAs, for various values of $r$, and $K=1,~p=0.08$. For $r=r^\ast=0.02987$, $g(K,n,p,r) \ge 0$; the minimum (zero) occurs at $n=n^\ast=33.98$ (see the text). The horizontal axis shows the number of years after age 70.}
\label{fig:Keq1}
\end{center}
\vspace{-20pt}
\end{figure}
As an example, behavior of $g(K,n,p,r)$ is shown in Fig.~\ref{fig:Keq1} for various values of $r$ and $K=1,~p=0.08$.
As seen in the figure, for given $K,p,r$, the equation $g(K,n,p,r)=0$
can have zero, one, or two solutions for $n$:
\begin{itemize}
\item No solutions (the solid green curve); here $g(K,n,p,r) > 0$ for all $n$, implying that the early-claimer is always ahead for this value of $r$, the average market return rate.
\item Only one solution, at $n=n^\ast$ for $r=r^\ast$ (the dotted red curve); here the beneficiary is always ahead except at $n=n^\ast$, where his position temporarily becomes neutral ($g=0$). 
The parameters $n^\ast,~r^\ast$ will play a critical role in our analysis, since $r>r^\ast$ implies $g(K,n,p,r) > g(K,n,p,r^\ast) \ge g(K,n^\ast,p,r^\ast)=0$ for all $n$.
\item Two distinct solutions, $n=n_1,~n_2$ (the dashed blue curve). For this value of $r$, the beneficiary starts falling behind at $n=n_1$ (age $70+n_1$) since $g < 0$ for $n_1~<~n~<~n_2$. The gain function is positive again for $n>n_2$; however, $70+n_2$ is too large a number to be relevant for most cases. For the parameters used in Fig.~\ref{fig:Keq1}, we have $n_1~=~20.87$ and $n_2~=~70.32$, again found numerically with Mathematica\cite{mathematica2022}.
\end{itemize}

\begin{table}[h!]
\centering
 \begin{tabular}{| c ||  c   c ||  c   c |  c  c |} 
 \hline
 $~K~$ & $~~n^\ast(q=0)~$ & $~~r^\ast(q=0)~$ &  $~~n^\ast(q=0.025)~$ & $~~r^\ast(q=0.025)~$ &   $~~n^\ast(q=0.037)~$ & $~~r^\ast(q=0.037)~$  \\ [0.5ex] 
 \hline
 1 & 33.98 & 0.02987 &  34.58 & 0.04394 & 35.69 & 0.05128 \\ 
 2 & 33.17 & 0.03061 &  33.53 & 0.04483 & 34.45 & 0.05221 \\
 3 & 32.39 & 0.03135 & 32.54  & 0.04573 & 33.30 & 0.05314 \\
 4 & 31.65 & 0.03210 & 31.62  & 0.04662 & 32.24 & 0.05406 \\
 5 & 30.93 & 0.03286 & 30.75  & 0.04751 & 31.26 & 0.05498 \\
 6 & 30.23 & 0.03363 & 29.93  & 0.04840 & 30.35 & 0.05589 \\
 7 & 29.57 & 0.03440 & 29.16  & 0.04928 & 29.52 & 0.05679 \\
 8 & 28.93 & 0.03517 & 28.44  & 0.05016 & 28.74 & 0.05767 \\
 \hline
\end{tabular}
\caption{\baselineskip 12pt The critical parameters $n^\ast$ and $r^\ast$ for the ``early claim with market gains'' scenario for various values of $K$, and $p=0.08$. The columns with $q=0$ (no COLAs) are solutions of Eqs.~\ref{eqn:nStar}, \ref{eqn:rStar}. The remaining columns with $q > 0$ (with COLAs) are solutions of Eqs.~\ref{eqn:nStarC}, \ref{eqn:rStarC}.}
\label{table:KnrTable}
\end{table}

In order to emphasize the meaning of these important parameters, we will go through an example. Making use of Table~\ref{table:KnrTable}, assume that a beneficiary claims his benefits at age 62 ($K=70-62=8$) and  invests his Social Security income in some market instrument (as detailed previously) with an average annual return rate of $r^\ast \simeq 3.5\%$. With these assumptions, his cumulative benefits will continue to exceed those from the ``late scenario,'' $T_E^M > T_L$, until he is $70+n^\ast\simeq 99$ years old. At that time his position becomes neutral ($T_E^M = T_L$) briefly (the break-even point from the point of view of the ``late scenario''), but he gets ahead again beyond that age. And a final crucial point is that, for any given $K$, if $r > r^\ast$ the beneficiary remains always ahead: $T_E^M > T_L$ for all $n$.

Next we examine how these results are modified when the cost-of-living adjustments are included in the analysis.

\subsection{``Early claim with market gains'' scenario--with COLAs} 
Making use of some of our earlier results, here we look at the effects of cost-of-living adjustments. With COLA's, the starting year benefit $S_K^C$ (see Eq.~\ref{eqn:SKC}) will grow both due to market gains and COLAs; thus, the total sum the beneficiary accumulates in the market before age 70 will now be
\begeqn
T_M^C = S_K^C\sum_{i=0}^{K-1}(1+q)^i(1+r)^i = \frac{S_0}{(1+p)^K(1+q)^K}\frac{(1+q)^K(1+r)^K-1}{(1+q)(1+r)-1}, \label{eqn:TMC}
\endeqn
which replaces $T_M$ of Eq.~\ref{eqn:TM}. (Recall that the COLA-modified terms have a superscript $C$). By  assumption, $T_M^C$ is left in the market after reaching 70 and continues to grow at the rate $r$. After 70, however, the new COLA-modified benefits are not invested but presumably used for other purposes. Thus, $T_E^M$ of Eq.~\ref{eqn:TEM} representing the accumulated benefits at age $70+n$ now becomes
\begeqnar
T_E^{MC} & = & T_M^C(1+r)^n + S_K\sum_{i=0}^{n-1}(1+q)^i, \text{~or} \nonumber \\
T_E^{MC} & = & \frac{S_0(1+r)^n}{(1+p)^K(1+q)^K}\frac{(1+q)^K(1+r)^K-1}{(1+q)(1+r)-1} + \frac{S_0}{(1+p)^K}\frac{(1+q)^n-1}{q}. \label{eqn:TEMC}
\endeqnar
The total benefits at age $70+n$ with the ``late scenario'' and including COLAs is still given by Eq.~\ref{eqn:TLC}:
\begeqn
T_L^C = S_0\frac{(1+q)^n-1}{q}. \label{eqn:TLC2}
\endeqn

At this point we can do a couple of consistency checks: Since $\lim_{q \to 0}[(1+q)^n-1]/q=n$,
$T_E^{MC}$ and $T_L^C$ reduce to $T_E^M$  and $T_L$ of Eqs.~\ref{eqn:TEM}, \ref{eqn:TL}, respectively, when the COLAs are ignored ($q\rightarrow 0$). Similarly, when the COLAs are retained but the market gains are ignored ($r \to 0$), $T_E^{MC}$ reduces to $T_E^C$ of Eq.~\ref{eqn:TEC} as expected.

As we saw earlier in Sect.~\ref{sect:noMwithC} on the effects of COLAs without the market gains, we intuitively expect that the benefits of claiming early will be reduced with the COLAs, even when the market gains are taken into account. In particular, we expect that the ratio of the average market return to the average COLA, $r/q$, will play an important role in determining how useful an early claim will be.  In order to understand these issues better, we again look at the gain function and its time derivative. The new gain function with the COLAs can be written as
\begeqnar
g^c(K,n,p,q,r) & \equiv & \frac{T_E^{MC}-T_L^C}{T_L^C}, \text{~~or} \label{eqn:gainMC} \\
g^c(K,n,p,q,r) & = & \frac{1}{(1+p)^K}\left\{\left[\frac{(1+q)^K(1+r)^K-1}{(1+q)^K[(1+q)(1+r)-1]}\right]\left[\frac{q(1+r)^n}{(1+q)^n-1}\right] +1 \right\} -1, \nonumber 
\endeqnar
which reduces to $g(K,n,p,r)$ of Eq.~\ref{eqn:gainM} in the limit $q \to 0$, as expected. Letting
\begeqn
A(K,p,q,r) \equiv  \frac{q}{(1+p)^K} \frac{(1+q)^K(1+r)^K-1}{(1+q)^K[(1+q)(1+r)-1]}, \label{eqn:A}
\endeqn
we can write the time derivative $\partial g^c/\partial n$ in the form
\begeqn
\frac{\partial g^c}{\partial n} = A(1+r)^n\left\{\frac{(1+q)^n \ln[(1+r)/(1+q)] - \ln(1+r)}{[(1+q)^n-1]^2}\right\}. \label{eqn:dgCdn}
\endeqn

\begin{figure}[htbp]
\vspace{-0pt}
\begin{center}
\includegraphics[width=4in]{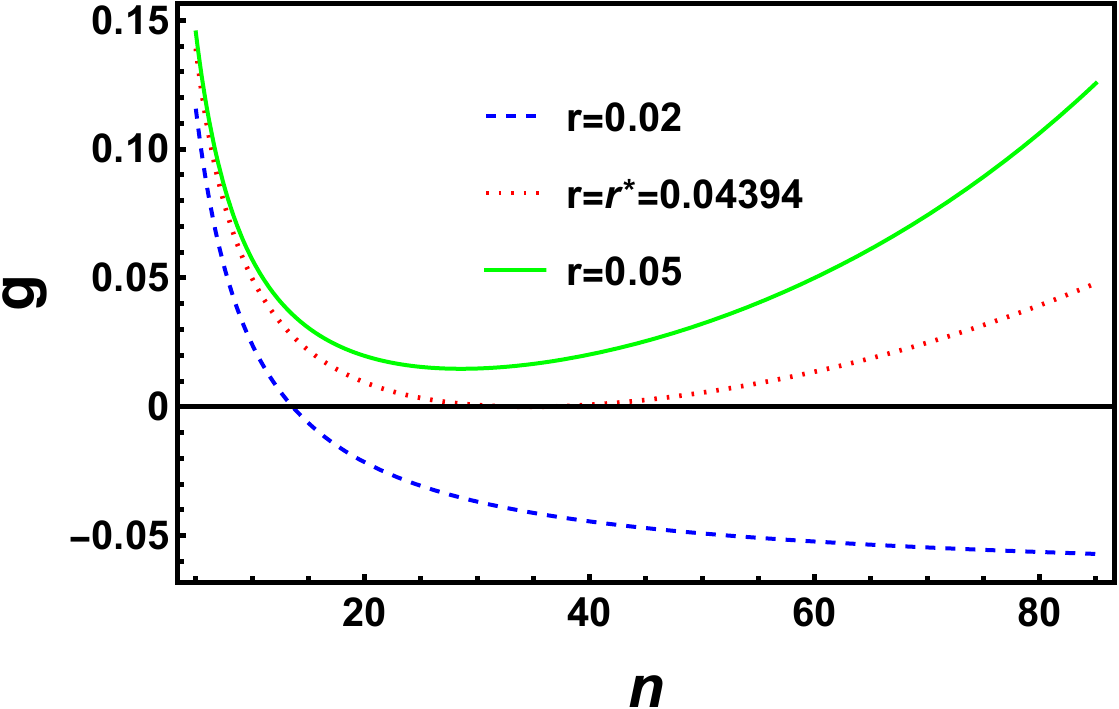}
\caption{\baselineskip 12pt The gain function $g^c(n)$ (with COLAs) for three different values of $r$, and $K=1,~p=0.08,~q=0.025$. For these parameters we have $r^\ast=0.04394,~n^\ast=34.58$ (Table~\ref{table:KnrTable}). The dashed blue curve: $r=0.02 < q$. The dotted red curve: $r=r^\ast > q$; it has a minimum at $n=n^\ast$ where $g^c(n^\ast,r^\ast)=0$. The solid green curve: $r=0.05 > r^\ast > q$.}
\label{fig:Keq1C}
\end{center}
\vspace{-20pt}
\end{figure}

By simple inspection, we can draw some general conclusions from Eqs.~\ref{eqn:gainMC}-\ref{eqn:dgCdn}:
\begin{itemize}
\item The coefficient $A(K,p,q,r) > 0$ for $r,q >0$. Thus,
\begeqn
\frac{\partial g^c}{\partial n} < 0 \text{~~for~} 0 < r \le q.
\endeqn
In other words, when the market return rate is less than the average COLA, the relative gain will be a monotonically decreasing function of time (for fixed $K,p,q,r$). In fact, for strict inequality, $r < q$, we have
\begeqn
\lim_{n \to \infty} g^c = \frac{1}{(1+p)^K} - 1 < 0.
\endeqn
Under these circumstances, the equation $g^c=0$ will have only one solution  ($n=n_1$). An early-claiming beneficiary will fall behind at age $70+n_1$ and will never recover ($g^c(K,n,p,q,r) \le 0$ for $n \ge n_1$).
\item For $r > q > 0$, $\lim_{n \to \infty} g^c = +\infty$, and the gain function displays a behavior similar to what we observed earlier when we considered the market gains without the COLAs. It will have a minimum at $n=n^\ast$, where $n^\ast$ is a solution of the equation $\partial g^c/\partial n = 0$. We can easily show that
\begeqn
n^\ast(q,r) = \frac{1}{\ln{(1+q)}}\ln\left\{\frac{\ln{(1+r)}}{\ln{[(1+r)/(1+q)]}} \right\}. \label{eqn:nStarC}
\endeqn
Note that $\lim_{r \to q}n^\ast = \infty$, consistent with the discussion above. Now the equation $g^c=0$ may again have zero, one or two solutions, depending on the parameters $K,p,q,r$. As in Sect.~\ref{sect:withMnoC}, the parameter $r^\ast$ will be a solution, for fixed $K,p,q$, of the equation
\begeqn
g^c[K,n^\ast(q,r^\ast),p,q,r^\ast]=0. \label{eqn:rStarC}
\endeqn
In order to re-emphasize the significance of the parameters $n^\ast,~r^\ast$, we summarize here some of their important properties:
\begeqnar
g_{min}(K,p,q,r) & = & g^c[K,n^\ast(q,r),p,q,r], \label{eqn:gminC} \\
g_{min}(K,p,q,r^\ast) & = & g^c[K,n^\ast(q,r^\ast),p,q,r^\ast] = 0. \label{eqn:gmin0}
\endeqnar
In other words, the gain function has a minimum at $n=n^\ast$, and that minimum is zero for $r=r^\ast$. A consequence of Eqs.~\ref{eqn:gCompare}, \ref{eqn:gminC}, \ref{eqn:gmin0}, for fixed $K,p,q$, is
\begeqn
g^c(K,n,p,q,r>r^\ast) > 0 \text{~~for~all~} n. \label{eqn:gGreater}
\endeqn
Thus, an early-claiming beneficiary will always be ahead if he can guarantee an average market return rate of $r > r^\ast$, even with the COLAs. However, we will find that, with everything else fixed, $r^\ast$ will be higher with the COLAs.
\end{itemize}
These different types of behavior for the gain function $g^c$ are illustrated in Fig.~\ref{fig:Keq1C} for $K=~1,~p=0.08$ and $q=0.025$, the average COLA for the years $1983-2022$. The dashed blue curve has $r=0.02 < q$; hence the negative slope and a zero at $n_1=13.6$. The early claimer falls behind at $70+n_1\simeq 84$ and never recovers. The dotted red curve is for $r=r^\ast=0.04394 > q$. It has a minimum (zero) at $n^\ast=34.58:~g^c(n^\ast,r^\ast)=0$. The solid green curve has $r=0.05 > r^\ast > q$; hence $g^c > 0$ for all $n$.

In Table~\ref{table:KnrTable}, columns with $q>0$ are solutions of the coupled Eqs.~\ref{eqn:nStarC} and \ref{eqn:rStarC}. Since they are both complicated functions of the parameters, they are solved iteratively for $\{n^\star, r\}$, with the converged solution yielding $\{n^\ast(q,r^\ast), r^\ast = r\}$. Note that the solutions of Eqs.~\ref{eqn:nStarC}, \ref{eqn:rStarC} ($q>0$, with COLA) will agree with those of Eqs.~\ref{eqn:nStar}, \ref{eqn:rStar} ($q=0$, no COLA) in the limit $q \to 0$. We find that we are able to reproduce the $q=0$ results in the Table (obtained using Eqs.~\ref{eqn:nStar}, \ref{eqn:rStar}) from Eqs.~\ref{eqn:nStarC}, \ref{eqn:rStarC} with $0 < q < 10^{-5}$. However, $q=0$ is a singular limit for the equations with COLA, since they are undefined at $q=0$.

Some general comments regarding the $\{n^\ast,r^\ast\}$ values in Table~\ref{table:KnrTable} may be helpful. Recall that the gain function (Eq.~\ref{eqn:gainMC}) has a minimum at $n=n^\ast(q,r)$. The minimum value is zero for $r=r^\ast$, i.e., $g^c(n^\ast,r^\ast)=0$, where we assume the other parameters, $K,p,q$ are held fixed. Recalling Eqs.~\ref{eqn:gminC}-\ref{eqn:gGreater}, for $r > r^\ast$ we have
\begeqn
g^c(n,r) \ge g^c(n^\ast,r) > g^c(n^\ast,r^\ast)=0.
\endeqn
Thus, a beneficiary who files early will have a positive gain throughout his life, if he can maintain an average market return rate of $r > r^\ast$. The values of $n^\ast$ in Table~\ref{table:KnrTable} are approximately in the range $28-36$; thus, the minimum of the gain function occurs when the beneficiary is in his late 90's $(70+28)$ at the earliest (for $r \ge r^\ast$). In order to maintain positive gains during this time, he has to ensure an average market return rate of approximately $3-3.5\%$ (depending on how early he claims), if the COLAs are not taken into account ($q=0$ columns in the table). With COLAs, the required rate of return is approximately $4.4-5.0\%$ for $q=2.5\%$, the average COLA for the years $1983-2022$. The required rate may be as high as $5.8\%$ if $q=3.7\%$, the average  for the years $1975-2022$ (the entire data set at \cite{cola2022}), is used.

\section{Optimization} \label{sect:optimize}
In the previous section we were concerned with finding parameter regimes where an early claimer can have positive gains with respect to the ``late scenario,'' possibly for the rest of his life, when both the cost-of-living adjustments and possible market gains are taken into account. It is possible to go a step further and find an optimal time for claiming benefits as a function of the remaining parameters in the problem. Mathematically, the problem reduces to finding a maximum for the gain function $g^c=g^c(K,n,p,q,r)$ in a 5-dimensional space. Clearly the problem is quite unwieldy in this form; by choosing $p=0.08$, as we have done throughout this work, and using $q=0.025$ for the average COLA parameter, the dimension can be reduced to a more manageable three. Then it is possible to find an optimal time for claiming benefits, or a $K_{opt}(n,r)$, as a function of an assumed average market return rate, $r$, and a time, $n$, in the future. Below we examine a couple of different optimization paths.

\subsection{Maximizing the minimum gain}
Here instead of choosing a specific time in the future, we let $n=n^\ast$ (see Eq.~\ref{eqn:nStarC}) where the gain function has its minimum, i.e., we seek a $K_{opt}$ that maximizes  $g^c_{min}$. Recall that for the range of parameters $q,~r$ we have been considering, $n^\ast\simeq 28-36$ (Table~\ref{table:KnrTable}), corresponding to an age range of 98-106.

The mathematical problem can now be stated as follows:
\begin{itemize}
\item Find $K=K_{opt}(n^\ast,r)$ that maximizes $g^c(K,n^\ast,p,q,r)$ for $p=0.08,~q=0.025$ and an arbitrary $r>q$ (Recall that $g^c$ is monotonically decreasing for $r\le q$, a regime we try to avoid).
\end{itemize}

\begin{figure}[htbp]
\vspace{-0pt}
\begin{center}
\includegraphics[width=3.3in]{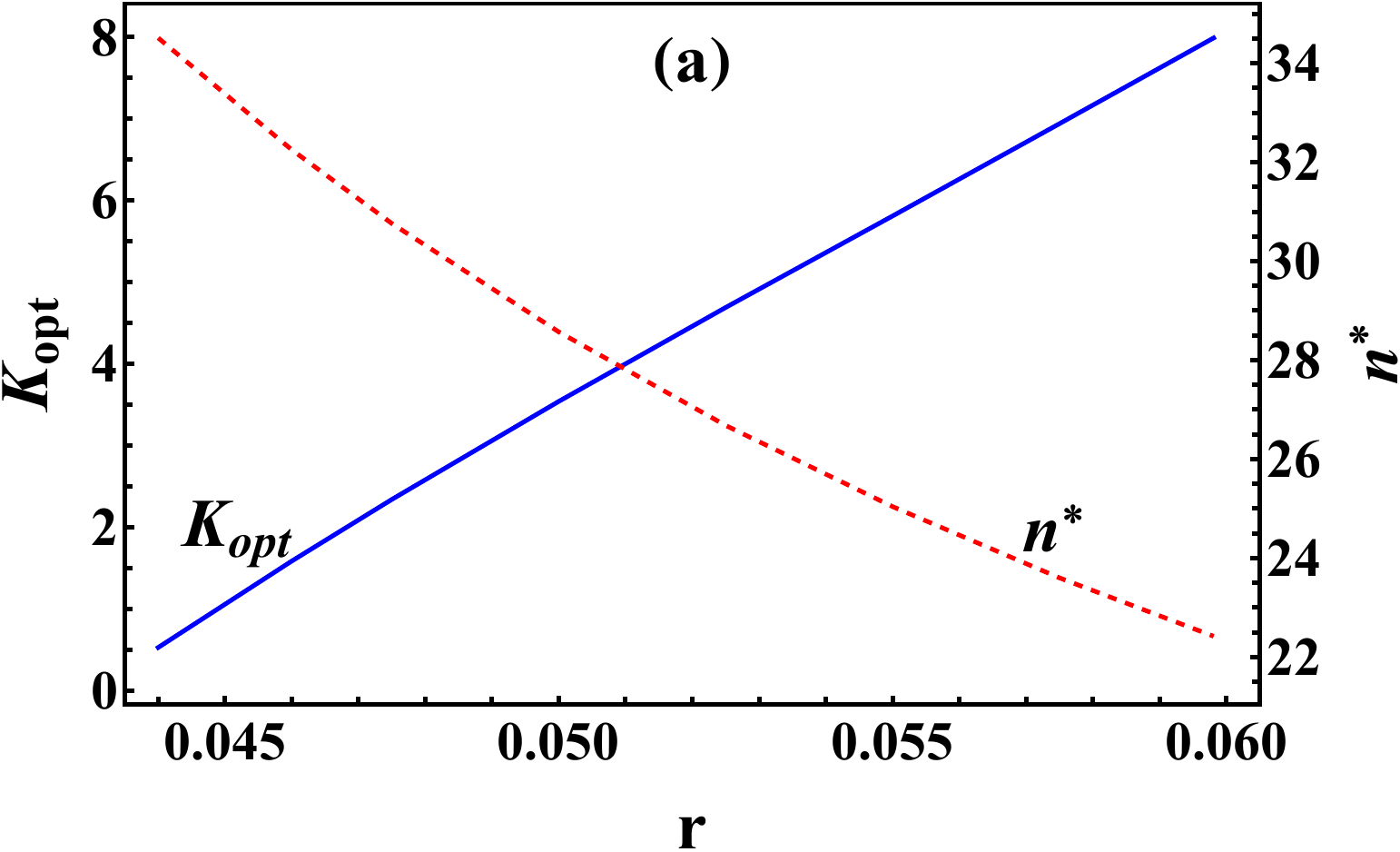}
\includegraphics[width=3.11in]{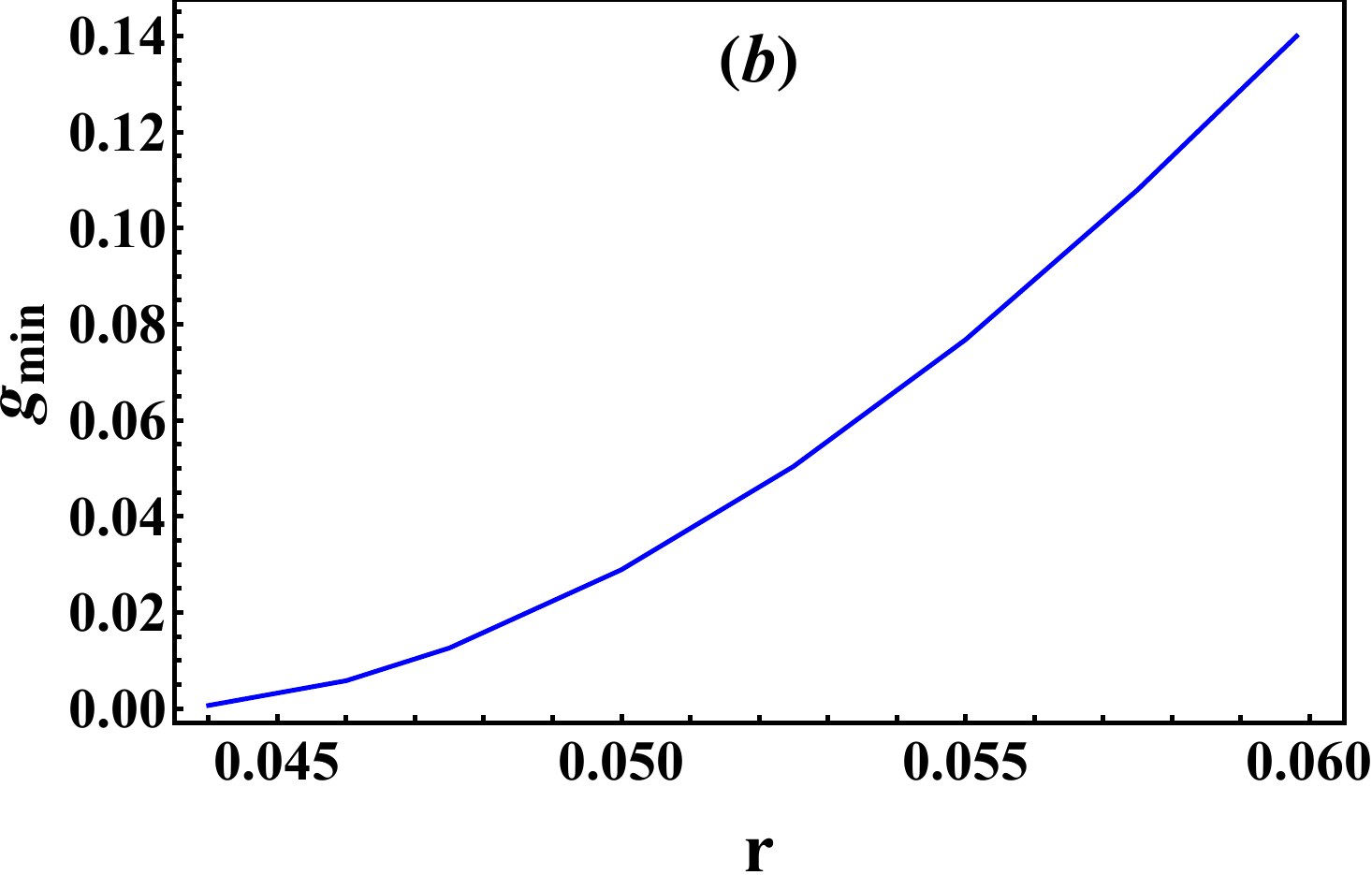}
\caption{\baselineskip 12pt (a) $K_{opt}$ that maximizes the minimum of the relative gain, $g^c_{min}$, as a function of the assumed average market return rate $r$. Recall that the optimal age is given by $70-K_{opt}$. Also shown is the parameter $n^\ast(q,r)$ where the minimum occurs. Again recall that $n$ measures years after age $70$; thus, $g^c_{min}$ occurs at age $70+n^\ast$. (b) $g^c_{min}$ as a function of $r$. }
\label{fig:Kopt_nStar}
\end{center}
\vspace{-20pt}
\end{figure}

We start by slightly rewriting $g^c$ of Eq.~\ref{eqn:gainMC}:
\begeqnar
g^c(K,n,p,q,r) & = & B(n,q,r)\left\{[(1+r)/(1+p)]^K - \frac{1}{[(1+p)(1+q)]^K} \right\} + \frac{1}{(1+p)^K} -1,  \nonumber \\
\text{where~~}B(n,q,r) & \equiv & \frac{q(1+r)^n}{[(1+q)(1+r)-1][(1+q)^n-1]}. \label{eqn:gainMCb}
\endeqnar
Then treating $K$ as a continuous variable and setting $\partial g^c/\partial K=0$ leads to
\begeqn
B\left\{(1+r)^K\ln\left(\frac{1+r}{1+p}\right) + \frac{\ln[(1+p)(1+q)]}{(1+q)^K}\right\} - \ln(1+p)=0. \label{eqn:dgdK}
\endeqn
Although Eq.~\ref{eqn:dgdK} can be solved for $K=K_{opt}$ analytically in certain limits (e.g., $r=p$), in general it requires  a numerical solution. Figure~\ref{fig:Kopt_nStar} shows $K_{opt}(n^\ast,r)$ (from Eq.~\ref{eqn:dgdK}), $n^\ast(q,r)$ (from Eq.~\ref{eqn:nStarC}) and $g^c_{min}=g^c(K_{opt},n^\ast,q,p,r)$ as functions of $r$. 
The limits for the average market return rate, $0.0440 \le r \le 0.0598$, are chosen to keep $K_{opt}$ within the allowed arrange of  $0 < K_{opt} \le 8$ (the optimal age has to be within $62-70$). A number of observations follows from Fig.~\ref{fig:Kopt_nStar}:
\begin{itemize}
\item $K_{opt}$ increases approximately linearly with the market return rate (Fig.~\ref{fig:Kopt_nStar}(a)); thus, the higher the market rate, the earlier is the optimal claiming age, $70-K_{opt}$, which is intuitively obvious.
\item The optimized minimum gain, $g^c_{min}$, increases faster than linearly with $r$   (Fig.~\ref{fig:Kopt_nStar}(b)). At $r=6\%$, the optimal claiming age is 62 ($K_{opt}=7.99$). The beneficiary's minimum gain with respect to the late scenario is $14\%$, which occurs at age $70+n^\ast=92.4$.  At any other age, his relative gain is higher since $g^c(K_{opt},n,p,q,r) > g_{min}$ for $n\ne n^\ast$.
\item The sum $K_{opt}+n^\ast$, representing the time span from the claiming age to when the gain function attains its minimum, is relatively constant: $30.4 \le  K_{opt}+n^\ast \le 35.0$, which has two implications:
\begin{itemize}
\item When calculating the average COLA rate $q$ (Eq.~\ref{eqn:qDef}) or the expected average market return rate $r$ (Eq.~\ref{eqn:rDef}), the time span to consider is approximately 30-35 years from the start of the benefits. This is a fairly long period that should be enough to smooth out the market or economic fluctuations. Thus $r=5-6\%$ and $q=2.5\%$ are not unreasonable assumptions.
\item Regardless of where the optimal claiming age happens to be, the optimized minimum gain is approximately 30-35 years in the future from that point. And neither ill-health nor extreme longevity will leave the beneficiary worse off than what is implied by $g^c_{min}$. Therefore, finding a claiming age that maximizes the minimum of the relative gain is a reasonable strategy.
\end{itemize}
\end{itemize}

\subsection{Optimizing for a particular age}

\begin{figure}[htbp]
\vspace{-0pt}
\begin{center}
\includegraphics[width=4.0in]{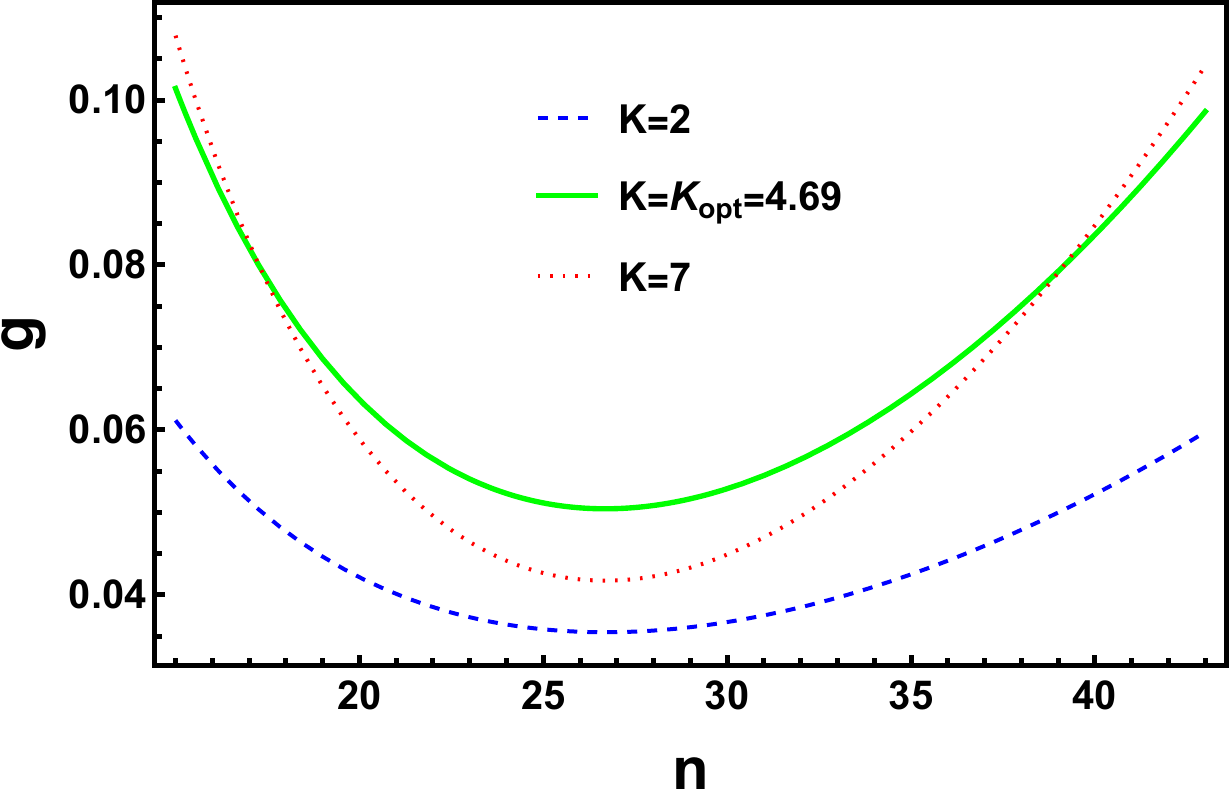}
\caption{\baselineskip 12pt The relative gain function $g^c(K,n,p,q,r)$ for three different values of $K$, with $p=0.08, q=0.025$ and $r=0.0525$. The solid green curve is for $K=K_{opt}$, optimized for $n=n^\ast$. The dashed blue curve for $K=2 < K_{opt}$ always remains below the optimized curve; however, the dotted red curve for $K=7>K_{opt}$ crosses above the optimal curve at $n=n_1=17.28$ and $n=n_2=39.13$.}
\label{fig:gOpt}
\end{center}
\vspace{-20pt}
\end{figure}

At this point a complication that we should consider is the following: Optimizing $g^c_{min}$ does not  imply that the gain function itself has been optimized for all ages. In other words, we may still have $g^c(K_{opt},n,p,q,r) < g^c(K,n,p,q,r)$ for some values of $K \ne K_{opt},~n \ne n^\ast$. In fact, the equation
\begeqn
g^c(K_{opt},n,p,q,r)=g^c(K,n,p,q,r),
\endeqn
with fixed $p,q,r$ tends to have two distinct solutions, $n=n_1,n_2$, for $K>K_{opt}$. Thus the gain function for $K>K_{opt}$ will be higher for $n<n_1$ and $n>n_2$.
This point is illustrated in Fig.~\ref{fig:gOpt} where we plot $g^c$ as a function of $n$ for various values of $K$, and $p=0.08,~q=0.025,~r=0.0525$. The minima for all three curves are at $n=n^\ast(q,r)=26.7$, and clearly the minimum for $K=K_{opt}=4.69$ curve is higher than those of the $K=2,7$ curves. However, the $K=7$ curve (dotted red) crosses above the $K=K_{opt}$ curve (solid green) at $n_1=17.28$ and $n_2=39.13$. In other words, although $g^c_{min}$ is optimized with $K=K_{opt}=4.69$, for the years corresponding to $n<n_1$ (earlier than age 87), or $n>n_2$ (later than age 109), the beneficiary may be better off with $K=7$, i.e., claiming at age 63 instead of $70-K_{opt}\simeq 65$. 

In order to address this particular complication, the gain function itself (not its minimum) can be optimized for any given age using the analytic expressions given above. The mathematical problem we need to solve now is the following:
\begin{itemize}
\item Find solutions of Eq.~\ref{eqn:dgdK} for $K=K_{opt}(n,r)$, which maximizes $g^c(K,n,p,q,r)$ with given $p,q,r$, and for various values of $n$ while ensuring that $0 < K_{opt} \le 8$. 
\end{itemize}

\begin{figure}[htbp]
\vspace{-0pt}
\begin{center}
\includegraphics[width=3.11in]{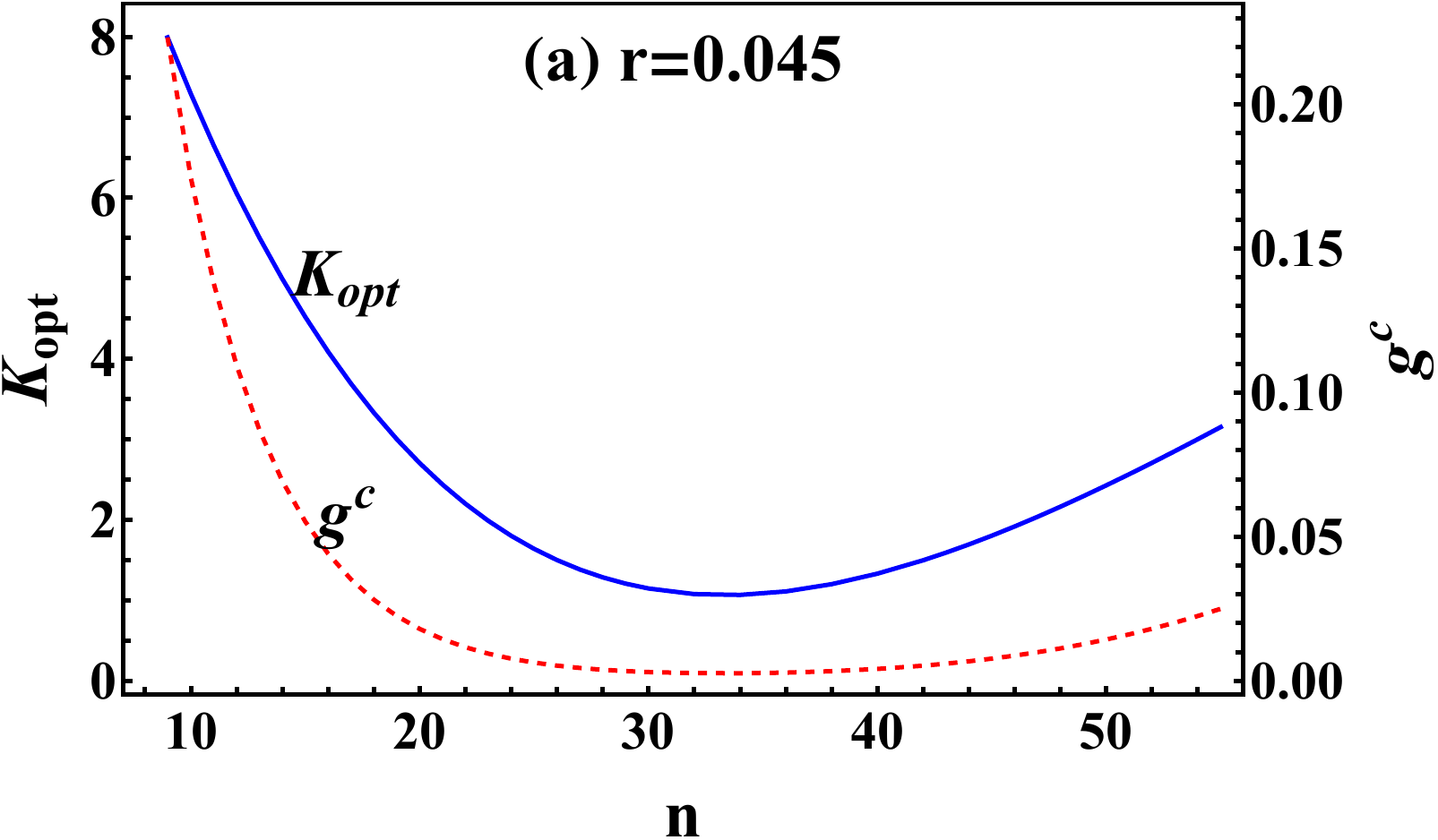}
\includegraphics[width=3.19in]{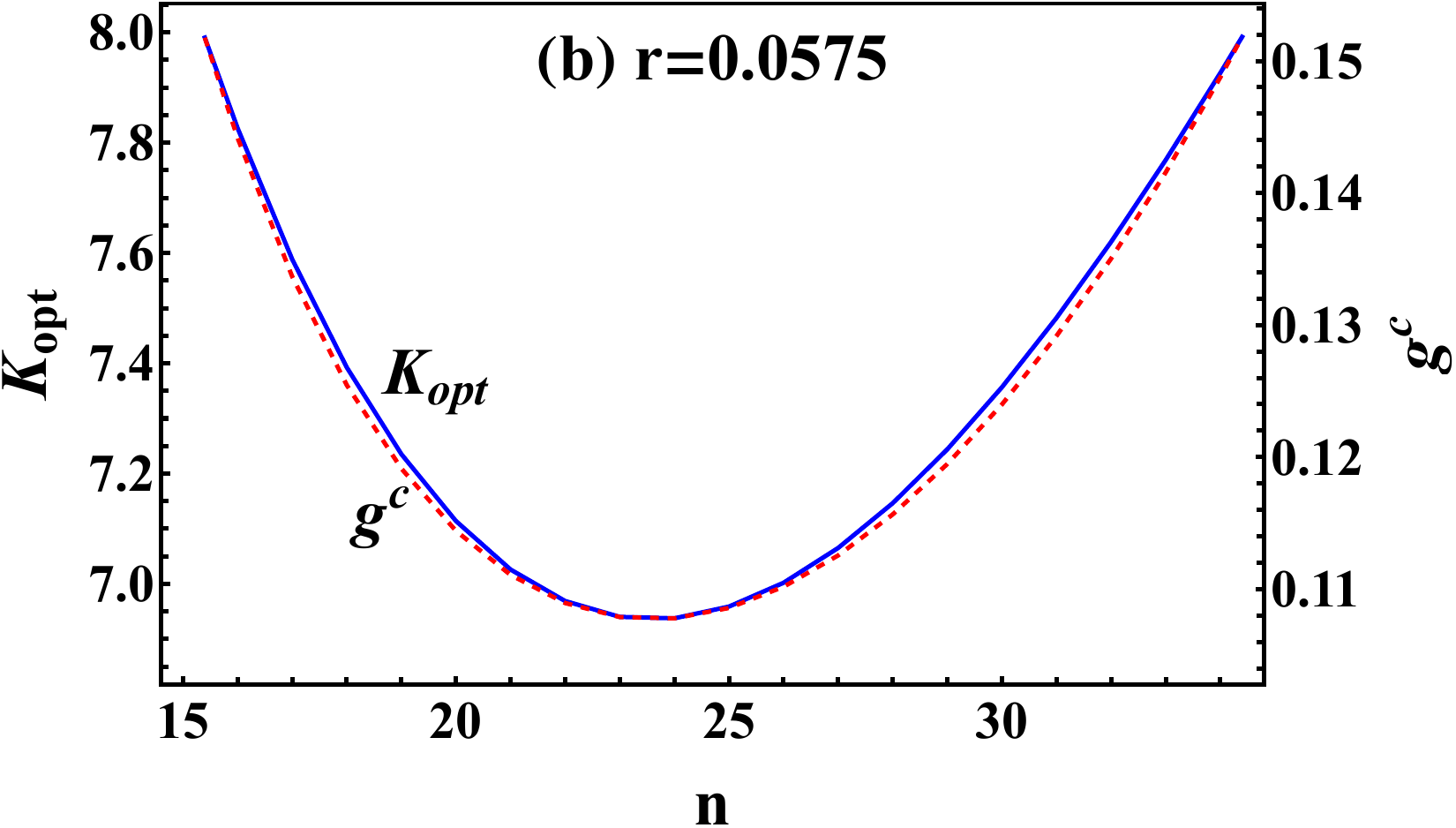}
\caption{\baselineskip 12pt $K_{opt}$ and the optimized value of the gain function $g^c$ as a function of $n$, the number of years after age 70. Recall that the optimal claiming age is given by $70-K_{opt}$, and $g^c$ measures the gain with respect to the ``late scenario'' where the beneficiary waits until age 70 to claim. (a) Average market return rate $r=0.045$. (b) $r=0.0575$. Note that in (b) although the two curves nearly coincide, the scales are different. }
\label{fig:Kopt-n-ab}
\end{center}
\vspace{-20pt}
\end{figure}

A couple of examples are shown in Fig.~\ref{fig:Kopt-n-ab}, again for $p=0.08,~q=0.025$. Recall that $g^c_{min}(K,p,q,r^\ast)=g^c(K,n^\ast,p,q,r^\ast)=0$. For the first example, in order to ensure $g^c_{min} > 0$, we choose $r=0.045$, which is just above the minimum $r^\ast$ for $q=0.025$ in Table~\ref{table:KnrTable}. The results are shown in Fig.~\ref{fig:Kopt-n-ab}(a). For any given $n$, the figure shows $K_{opt}$ and the resulting optimized gain. For example, for $n=10$ (age 80), we have $K_{opt}=7.29,~g^c=0.175$, a again of 17.5\% over the late scenario.
For $n=20$ (age 90), we get $K_{opt}=2.70$ and $g^c=0.018$, a gain of only $1.8\%$ over the late scenario. The minimum gain is at $n=n^\ast=33.34$ (age 103), where it is a mere $0.3\%$. In Fig.~\ref{fig:Kopt-n-ab}(b) where we assume a slightly higher return rate of $5.75\%$, the gains are markedly higher. For  $15 < n < 35$ (ages 85-105), the optimized $g^c$ varies between a minimum of $10.8\%$ at $n=n^\ast=23.6$ and $15.2\%$ at the end points of the range. Similarly, $K_{opt}$ varies little: $6.94 \le K_{opt} \le 8$ for this value of $r$. Thus, with a rather modest market return of $ 5.8\%$, a beneficiary who claims at age 62-63 would indeed be well ahead for the rest of his life.

\section{Summary and conclusions} \label{sect:summary}
The Social Security benefits provide a financial lifeline to most retirees, and a retiree typically claims his benefits as soon as he reaches his full-retirement age (FRA), or as early as age 62 if he is willing to accept some penalties. If a retiree is not financially constrained, however, there is an incentive for waiting until age 70 because of the built-in yearly increases to benefits (presently $8\%$) between his FRA and age 70. Ironically, precisely in those cases where filing at FRA or earlier is not a financial necessity, it may be beneficial to file early without waiting until age 70. This work focused on the decision of whether to do so and when the optimal time might be.

Starting with a simple ``break-even'' analysis that retirees typically do to find out when they would break even if they delayed retirement beyond FRA, we developed a series of successively more comprehensive analytic models that compare early-claiming scenarios with the ``late scenario'' where the beneficiary waits until age 70. First improvement on the break-even analysis was the inclusion of inflation through the cost-of-living adjustments (COLAs) in Sect.~\ref{sect:breakEven}. From the point of view of an early-claimer (a view adopted throughout this work) COLAs make early claims less attractive. Conversely, from the point of view of a late claimer, they bring the break-even point closer (Table~\ref{table:Kn1CTable}). 
Results of filing early and investing the benefits received in the financial markets were examined in Sect.~\ref{sect:marketGains}. As summarized in Table~\ref{table:KnrTable}, if the COLAs are ignored, even a modest market return of $3.5\%$ would be sufficient for a beneficiary who files at age 62 to be ahead of the late scenario all his life. If the COLAs are taken into account,  that return rate would have to be $5\%$ if the average COLA is $2.5\%$, and $5.8\%$ if the COLA rate is at its historical average of $3.7\%$. Since the timespan relevant to our discussion is approximately three decades from the time of claiming benefits, these market rates, averaged over 30-35 years, are not unreasonable.

If the beneficiary wants to be not just ahead of the late scenario but to optimize his gains, then he has a number of options, of which we examined only two. First, we found an optimal filing age that maximizes his minimum gain, $g^c_{min}$, as a function of the assumed market return rate (Fig.~\ref{fig:Kopt_nStar}). For example, if the average market rate $r=5.8\%$, we find $K_{opt}=6.9$, implying an optimal age of approximately 63 ($70-K_{opt}$) for claiming; the minimum gain of $11\%$ occurs at $n^\ast=23.6$, corresponding to an age of over 93 ($70+n^\ast$). Both before and after that age, the gain is higher (on either side of the minimum).

The second option we examined was finding an optimal claiming age that maximizes the relative gain at a specific point in time, for two different market rates (Fig.~\ref{fig:Kopt-n-ab}). We found that the claiming age is rather insensitive to the age for which the gain is maximized ($K_{opt}\simeq 7-8$) if $r=5.8\%$, although there is considerable variation in both $K_{opt}$ and the maximized gain for $r=4.5\%$.

The short conclusion from these analyses is that, under the conditions explained at the beginning of Sect.~\ref{sect:marketGains}, there is no financial incentive for waiting until age 70 for claiming Social Security benefits. A beneficiary would be almost always ahead if he were to claim quite early (age 62-63), if the market return rate on his investments over the next 30-35 years is approximately $5-6\%$. 

Real-life scenarios are usually more complex than the simplified mathematical models discussed in this work. We made some assumptions in order to make analytical progress, and we did not consider important factors such as taxes and spousal benefits. It would also be beneficial to examine optimization that takes into account a probabilistic survival function. We hope to address these issues in a future publication. It is important to note that, while we provided examples in our figures and tables for a variety of parameters, the analytic expressions derived here can be useful more generally. An individual can customize the choice of parameters to his own circumstances and use these models to make a timing decision that aligns with his goals better.

\acknowledgments
The author gratefully acknowledges useful comments by Gary Hallock that helped improve the manuscript.
This research did not receive any specific grant from funding agencies in the public, commercial, or
not-for-profit sectors.


\begin{thebibliography}{9}
\expandafter\ifx\csname natexlab\endcsname\relax\def\natexlab#1{#1}\fi
\expandafter\ifx\csname bibnamefont\endcsname\relax
  \def\bibnamefont#1{#1}\fi
\expandafter\ifx\csname bibfnamefont\endcsname\relax
  \def\bibfnamefont#1{#1}\fi
\expandafter\ifx\csname citenamefont\endcsname\relax
  \def\citenamefont#1{#1}\fi
\expandafter\ifx\csname url\endcsname\relax
  \def\url#1{\texttt{#1}}\fi
\expandafter\ifx\csname urlprefix\endcsname\relax\def\urlprefix{URL }\fi
\providecommand{\bibinfo}[2]{#2}
\providecommand{\eprint}[2][]{\url{#2}}

\bibitem[{\citenamefont{Coile et~al.}(2002)\citenamefont{Coile, Diamond,
  Gruber, and Jousten}}]{coile2002}
\bibinfo{author}{\bibfnamefont{C.}~\bibnamefont{Coile}},
  \bibinfo{author}{\bibfnamefont{P.}~\bibnamefont{Diamond}},
  \bibinfo{author}{\bibfnamefont{J.}~\bibnamefont{Gruber}}, \bibnamefont{and}
  \bibinfo{author}{\bibfnamefont{A.}~\bibnamefont{Jousten}},
  \bibinfo{journal}{Journal of Public Economics} \textbf{\bibinfo{volume}{84}},
  \bibinfo{pages}{357} (\bibinfo{year}{2002}).

\bibitem[{\citenamefont{Friedman and Phillips}(2008)}]{friedman2008}
\bibinfo{author}{\bibfnamefont{J.}~\bibnamefont{Friedman}} \bibnamefont{and}
  \bibinfo{author}{\bibfnamefont{H.}~\bibnamefont{Phillips}},
  \bibinfo{journal}{Financial Services Review} \textbf{\bibinfo{volume}{17}},
  \bibinfo{pages}{155} (\bibinfo{year}{2008}).

\bibitem[{\citenamefont{Alleva}(2016)}]{alleva2016}
\bibinfo{author}{\bibfnamefont{B.~J.} \bibnamefont{Alleva}},
  \bibinfo{journal}{Social Security Bulletin} \textbf{\bibinfo{volume}{76}},
  \bibinfo{pages}{1} (\bibinfo{year}{2016}).

\bibitem[{\citenamefont{Ali et~al.}(2019)\citenamefont{Ali, Fang, Sota, Taylor,
  and Wang}}]{ali2019}
\bibinfo{author}{\bibfnamefont{Y.}~\bibnamefont{Ali}},
  \bibinfo{author}{\bibfnamefont{M.}~\bibnamefont{Fang}},
  \bibinfo{author}{\bibfnamefont{P.~A.~A.} \bibnamefont{Sota}},
  \bibinfo{author}{\bibfnamefont{S.}~\bibnamefont{Taylor}}, \bibnamefont{and}
  \bibinfo{author}{\bibfnamefont{X.}~\bibnamefont{Wang}},
  \bibinfo{journal}{Risks} \textbf{\bibinfo{volume}{7}}, \bibinfo{pages}{124}
  (\bibinfo{year}{2019}).

\bibitem[{\citenamefont{Diamond et~al.}(2021)\citenamefont{Diamond, Boyd,
  Greenberg, Kochenderfer, and Ang}}]{diamond2021}
\bibinfo{author}{\bibfnamefont{S.}~\bibnamefont{Diamond}},
  \bibinfo{author}{\bibfnamefont{S.}~\bibnamefont{Boyd}},
  \bibinfo{author}{\bibfnamefont{D.}~\bibnamefont{Greenberg}},
  \bibinfo{author}{\bibfnamefont{M.}~\bibnamefont{Kochenderfer}},
  \bibnamefont{and} \bibinfo{author}{\bibfnamefont{A.}~\bibnamefont{Ang}},
  \bibinfo{journal}{\url{https://doi.org/10.48550/arXiv.2106.00125}}
  (\bibinfo{year}{2021}).

\bibitem[{\citenamefont{\url{https://www.ssa.gov/benefits/retirement/planner/agereduction.html}.}()}]{fra2022}
\bibinfo{author}{\bibnamefont{\url{https://www.ssa.gov/benefits/retirement/planner/agereduction.html}.}}

\bibitem[{\citenamefont{\url{https://www.ssa.gov/benefits/retirement/planner/delayret.html}.}()}]{delayedRet2022}
\bibinfo{author}{\bibnamefont{\url{https://www.ssa.gov/benefits/retirement/planner/delayret.html}.}}

\bibitem[{\citenamefont{\url{https://www.ssa.gov/oact/cola/colaseries.html}.}()}]{cola2022}
\bibinfo{author}{\bibnamefont{\url{https://www.ssa.gov/oact/cola/colaseries.html}.}}

\bibitem[{\citenamefont{{Wolfram Research, Inc.}}()}]{mathematica2022}
\bibinfo{author}{\bibnamefont{{Wolfram Research, Inc.}}},
  \emph{\bibinfo{title}{Mathematica, {V}ersion 13.1}},
  \bibinfo{note}{{C}hampaign, IL, 2022},
  \urlprefix\url{https://www.wolfram.com/mathematica}.

\end{thebibliography}

\end{document}